\documentclass[12pt]{article}
\title{
\vspace{-3mm}
\rightline{\small IFUP-TH 2002/2}
\vspace{8mm}
\bf Accounting for the finiteness of the Higgs-boson mass in the 3D
Georgi-Glashow model}
\author{Dmitri Antonov \thanks{
E-mail address: {\tt antonov@df.unipi.it}}
\\
{\it INFN-Sezione di Pisa, Universit\'a degli studi di Pisa,}\\
{\it Dipartimento di Fisica, Via Buonarroti, 2 - Ed. B -
I-56127 Pisa, Italy}\\
{\it and}\\
{\it Institute of Theoretical and Experimental Physics,}\\
{\it B. Cheremushkinskaya 25, RU-117 218 Moscow, Russia}}
\date{}
\begin{document}
\maketitle
\vspace{1mm}
\centerline{\bf {Abstract}}
\vspace{3mm}
\noindent
(2+1)-dimensional Georgi-Glashow model is explored in the
regime when the Higgs boson is not infinitely heavy, but its mass
is rather of the same order of magnitude as the mass of the W boson.
In the weak-coupling limit, the Debye mass of the
dual photon and the expression for the monopole potential are found.
The cumulant expansion applied to the average over the Higgs field
is checked to be convergent for the known data on the monopole
fugacity. These results are further generalized to the $SU(N)$-case.
In particular, it is found that the requirement of convergence of the cumulant expansion
establishes a certain upper bound on the number of colours.
This bound, expressed in terms of the parameter of the weak-coupling
approximation, allows the number of colours to be large enough. Finally, the
string tension and the coupling constant of the so-called rigidity term
of the confining string are found at arbitrary number of colours.

\vspace{10mm}

\section{Introduction}

Since the second half of the seventies~\cite{1},
(2+1)-dimensional Georgi-Glashow model is known as
an example of the theory allowing for an analytic description
of confinement. However, confinement in the Georgi-Glashow model is typically discussed
in the limit of infinitely large Higgs-boson mass, when the
model is reduced to compact QED.
In ref.~\cite{nd}, possible influence of the Higgs field to the dynamics of the Georgi-Glashow
model has been studied both at zero and nonzero temperatures. This has been done
not only under the assumption that the Higgs-field mass is finite rather than infinite,
which allows this field to propagate, but
in the Bogomolny-Prasad-Sommerfield (BPS) limit~\cite{bps}. This is the limit
when the Higgs field is {\it much} lighter than the W boson (but is still much heavier
than the dual photon).
The first aim of the present paper is to generalize the zero-temperature results of
ref.~\cite{nd}
to the case when the mass of the Higgs boson is of the same order of magnitude
as the mass of the W boson. This situation is thus intermediate between the BPS limit
and the limit of compact QED. In this way, we shall find the monopole potential and the Debye mass,
and prove the convergence of the cumulant expansion associated to the average over the Higgs field.
This will be done in the next Section.

Another aim of the present paper, which will be realized in Section~3, is to generalize this analysis to the
$SU(N)$-case. The Debye mass and the parameter of the
cumulant expansion will then be $N$-dependent quantities. The $N$-dependence of the
latter will yield a certain upper bound on $N$ necessary to ensure the convergence
of the cumulant expansion. This bound will turn out to be the exponent of the inverse parameter of the
weak-coupling approximation, that will allow $N$ to vary in a wide enough range. We shall also
find the values of the two leading coupling constants of the confining-string Lagrangian at arbitrary $N$.

The main results of the paper will be summarized in the Conclusions.
In the Appendix, some technical details of the performed calculations will finally be outlined.

\section{SU(2)-case}

The Euclidean action of the (2+1)D Georgi-Glashow
model reads~\cite{1}

\begin{equation}
\label{GG}
S=\int d^3x\left[\frac{1}{4g^2}\left(F_{\mu\nu}^a\right)^2+
\frac12\left(D_\mu\Phi^a\right)^2+\frac{\lambda}{4}\left(
\left(\Phi^a\right)^2-\eta^2\right)^2\right].
\end{equation}
Here, the Higgs field $\Phi^a$ transforms by the adjoint representation and
$D_\mu\Phi^a\equiv\partial_\mu\Phi^a+\varepsilon^{abc}A_\mu^b
\Phi^c$. Next, $\lambda$ is the Higgs coupling constant of dimensionality [mass],
$\eta$ is the Higgs {\it v.e.v.} of dimensionality $[{\rm mass}]^{1/2}$, and
$g$ is the electric coupling constant of the same dimensionality.

At the one-loop level, the partition function of the theory~(\ref{GG}) takes
the following form~\cite{dietz}:

\begin{equation}
\label{1}
S=\int d^3x\left[\frac12(\nabla\chi)^2+\frac12(\nabla\psi)^2
+\frac{m_H}{2}\psi^2-2\zeta{\rm e}^{g_m\psi}\cos(g_m\chi)\right].
\end{equation}
Here, $\chi$ is the dual-photon field, and the field $\psi$ accounts
for the Higgs field, when it is not infinitely heavy
({\it i.e.} one deviates from
the compact-QED limit). Next, $g_m$ is the magnetic coupling constant
related to the electric one as
$g_mg=4\pi$. The Higgs-boson mass, $m_H$, reads $m_H=\eta\sqrt{2\lambda}$, and
the monopole fugacity $\zeta$ has the form:

\begin{equation}
\label{Ze}
\zeta=\frac{m_W^{7/2}}{g}\delta\left(\frac{\lambda}{g^2}\right)
{\rm e}^{-(4\pi/g^2)m_W\epsilon}.
\end{equation}
In this formula, $m_W=g\eta$ stands for the W-boson mass,
and $\epsilon=\epsilon(\lambda/g^2)$ is a certain monotonic, slowly
varying function, $\epsilon\ge 1$, $\epsilon(0)=1$~\cite{bps},
$\epsilon(\infty)\simeq 1.787$~\cite{kirk}.
As far as the function $\delta$ is concerned,
it is determined by the loop corrections.
It is known~\cite{ks} that this function grows
in the vicinity of the origin ({\it i.e.} in the BPS limit). However,
the speed of this growth is so that it does not spoil the exponential smallness
of $\zeta$ in the standard weak-coupling regime $g^2\ll m_W$ (or $g\ll\eta$) which
we adapt in this paper.

Integrating further in eq.~(\ref{1}) over $\psi$ by virtue of the
cumulant expansion, we get:

\begin{equation}
\label{2}
S\simeq\int d^3x\left[\frac12(\nabla\chi)^2-2\xi\cos(g_m\chi)\right]-
2\xi^2\int d^3xd^3y\cos(g_m\chi({\bf x})){\cal K}({\bf x}-{\bf y})
\cos(g_m\chi({\bf y})).
\end{equation}
In this expression, we have disregarded all the cumulants of the orders higher
than the second, and the limits of applicability of
this so-called bilocal approximation will be discussed below.
In eq.~(\ref{2}), ${\cal K}({\bf x})\equiv {\rm e}^{g_m^2D_{m_H}({\bf
x})}-1$ with $D_{m_H}({\bf x})\equiv{\rm e}^{-m_H|{\bf x}|}/(4\pi|{\bf
x}|)$ standing for the Higgs-field propagator, and

\begin{equation}
\label{fug}
\xi\equiv\zeta{\rm e}^{\frac{g_m^2}{2}D_{m_H}(0)}=\frac{m_W^{7/2}}{g}
\delta\left(\frac{\lambda}{g^2}\right){\rm e}^{\frac{2\pi m_W}{g^2}
\left(-2\epsilon+{\rm e}^{-c}\right)}
\end{equation}
denotes the modified fugacity.
In the derivation of eq.~(\ref{fug}), we have in the standard way set
$m_W$ for the UV cutoff in the weak-coupling regime and denoted
$c\equiv m_H/m_W$.

As it is clear
from eq.~(\ref{2}), the compact-QED limit
is achieved when $m_H$ formally tends to infinity, {\it i.e.} $c\to\infty$.
In ref.~\cite{nd}, there has been explored the opposite, BPS, limit $c\ll 1$.
Since $D_{m_H}({\bf x}-{\bf y})\sim m_H$, one can impose the
inequality $g_m^2m_H\ll 1$, which together with the weak-coupling
approximation yields $c\ll 1$, and obtain from eq.~(\ref{2}) the
following action:

$$
S\simeq\int d^3x\left[\frac12(\nabla\chi)^2-2\xi\cos(g_m\chi)\right]-
2(g_m\xi)^2\int d^3xd^3y\cos(g_m\chi({\bf x}))D_{m_H}({\bf x}-{\bf y})
\cos(g_m\chi({\bf y})).
$$
Note that according to eq.~(\ref{fug}), the modified fugacity
$\xi$ remains to be
exponentially small in this limit. That is firstly because
$\epsilon>\frac{{\rm e}^{-c}}{2}\simeq\frac12$ and secondly because, as
it was discussed above, according to ref.~\cite{ks}, the function
$\delta$ entering eq.~(\ref{Ze}) grows at $c\ll 1$ slower than
exponentially. Next, the fact that $D_{m_H}({\bf x})$ rapidly vanishes at
$|{\bf x}|\to\infty$ enables one to estimate the
parameter of the cumulant expansion, which in this case reads
$\xi g_m^2\int d^3xD_{m_H}({\bf x})=g_m^2\xi/m_H^2$. This quantity
is exponentially small due to the exponential smallness of $\xi$,
which proves the convergence of the cumulant expansion.

In what follows, we shall explore the action~(\ref{2})
in the regime intermediate between the BPS- and compact-QED limits,
namely $c\sim 1$. First of all note that since $c^2=2\lambda/g^2$,
$\xi$ will be exponentially small provided that $\epsilon(x)>
{\rm e}^{-\sqrt{2x}}/2$ at $x\sim 1/2$. One can see that this inequality is
always satisfied, since its r.h.s.
is not larger than 1/2, while $\epsilon\ge 1$.
Next, analogously to the case $c\ll 1$,
by noting that ${\cal K}({\bf x})$ rapidly
vanishes at $|{\bf x}|\to\infty$, the parameter of the
cumulant expansion can be estimated as $\xi I$, where
$I\equiv\int d^3x{\cal K}({\bf x})$. This integral is
evaluated in the Appendix. At $\frac{a}{c}{\rm e}^{-c}\gg 1$ with
$a\equiv4\pi m_H/g^2$ (which is obviously true in the weak-coupling regime),
it reads

\begin{equation}
\label{3}
I\simeq\frac{4\pi}{m_H^3}\left\{\sum\limits_{n=1}^{[1/c]}
\frac{a^n}{nn!}\left[n^{n-2}\Gamma(3-n,cn)-c^{2-n}{\rm e}^{-cn}\right]
+c^2\left[\exp\left(\frac{a}{c}{\rm e}^{-c}\right)\left(1-\frac{c}{a}{\rm e}^c\right)+
({\rm e}-1)\ln a\right]\right\}.
\end{equation}
Here, $[1/c]$ stands for the largest integer, smaller or equal to
$1/c$, and $\Gamma(b,x)=\int\limits_{x}^{\infty}
dt{\rm e}^{-t}t^{b-1}$ denotes the incomplete Gamma-function.
In the case $c\sim 1$ under study, the sum entering eq.~(\ref{3}) contains
a few terms, among whose the dominant one is of the order of $a$.
These terms can thus be disregarded with respect to the term of the order of ${\rm e}^a$
standing in that equation, and we finally obtain:
$I\simeq\frac{4\pi}{m_Hm_W^2}\exp\left(\frac{a}{c}{\rm e}^{-c}\right)$.
Consequently, the parameter of the cumulant expansion, $\xi I$,
will be exponentially small, provided that $\epsilon(x)>\frac32
{\rm e}^{-\sqrt{2x}}$ at $x\sim 1/2$. In particular, we should have
$\epsilon(1/2)>3/(2{\rm e})\simeq 0.552$, which is clearly true,
since $\epsilon\ge 1$. Thus, cumulant expansion is convergent
in the case $c\sim 1$ under study.

One can further straightforwardly read off
from eq.~(\ref{2}) the squared Debye mass of the dual photon.
It has the form $m_D^2=2g_m^2\xi(1+2\xi I)$, where as it was just discussed,
the second term in the brackets is exponentially small with respect
to the first one, and therefore $m_D=g_m\sqrt{2\xi}(1+\xi I)$.
Obviously, unity and $\xi I$ here are the contributions to $m_D$ brought about
by the first and the second cumulants in eq.~(\ref{2}), respectively.
Note also that this result for $m_D$
obviously reproduces the compact-QED one (see {\it e.g.}~\cite{nd}),
$g_m\sqrt{2\zeta}$. Indeed, at $m_H\to\infty$, $\xi\to\zeta$ and,
as it follows
directly from the definition of $I$, $I\to 0$, that proves our statement.

Similarly to how it was done for the case $c\ll 1$ in ref.~\cite{nd},
it is also possible in our case $c\sim 1$ to derive the representation
of the action~(\ref{2}) in terms of dynamical monopole densities $\rho$'s.
To this end, one should perform in the partition function the following
substitution

$$\exp\left[-\frac12\int d^3x(\nabla\chi)^2\right]=\int {\cal D}\rho
\exp\left[-\frac{g_m^2}{2}\int d^3xd^3y\rho({\bf x})D_0({\bf x}-{\bf y})
\rho({\bf y})-ig_m\int d^3x\chi\rho\right],$$
where $D_0({\bf x})\equiv 1/(4\pi|{\bf x}|)$ is the Coulomb propagator.
After that, it is necessary to solve the resulting saddle-point equation

\begin{equation}
\label{sp}
\sinh(\phi({\bf x}))\left[1+2\xi\int d^3y{\cal K}({\bf x}-{\bf y})
\cosh(\phi({\bf y}))\right]=\frac{\rho({\bf x})}{2\xi},
\end{equation}
where $\phi\equiv ig_m\chi$. This equation can be solved iteratively by
imposing the Ansatz $\phi=\phi_1+\phi_2$ with $|\phi_2|\ll|\phi_1|$.
Introducing the notation $f\equiv\sqrt{1+\left(\frac{\rho}{2\xi}\right)^2}$, we then
obtain:

$$\phi_1({\bf x})={\rm arcsinh}\left(\frac{\rho({\bf x})}{2\xi}\right),~~
\phi_2({\bf x})=-\frac{\rho({\bf x})}{f({\bf x})}\int d^3y{\cal K}({\bf x}-{\bf y})
f({\bf y}).$$
On the other hand, the average monopole density stemming from eq.~(\ref{2})
reads $\frac{\partial\ln\int {\cal D}\chi{\rm e}^{-S}}{{\cal V}\partial\ln\xi}
\simeq 2\xi(1+2\xi I)$,
where ${\cal V}$ is the 3D-volume of observation.
Therefore, at $|\rho|\le\xi$, we have $f\sim 1$, $|\phi_1|\sim|\rho|/\xi$,
and $|\phi_2|\sim\xi I |\phi_1|\ll |\phi_1|$ thus justifying our Ansatz.
The obtained solution to
the saddle-point equation yields the representation of the theory~(\ref{2})
in terms of $\rho$'s in the form

\begin{equation}
\label{Act}
S=\frac{g_m^2}{2}\int d^3xd^3y\rho({\bf x})D_0({\bf x}-{\bf y})
\rho({\bf y})+V[\rho].
\end{equation}
Here, the multivalued potential of monopole densities reads

$$V[\rho]=\int d^3x\left[\rho{\,}{\rm arcsinh}\left(\frac{\rho}{2\xi}\right)
-2\xi f\right]-2\xi^2\int d^3xd^3yf({\bf x}){\cal K}({\bf x}-{\bf y})f({\bf y}).$$
Note that the multivaluedness of this potential realizes the world-sheet
independence of the Wilson loop in the theory~(\ref{2}). This is the essence of the
string representation of the Georgi-Glashow model, discussed for the compact-QED
limit, $c\to\infty$, in ref.~\cite{cs} and for the BPS-limit, $c\ll 1$, in
ref.~\cite{nd}.

Note also that at very low densities, $|\rho|\ll\xi$, up to an inessential constant addendum,
$V[\rho]\simeq\frac{g_m^2}{2m_D^2}
\int d^3x\rho^2$, {\it i.e.} the action~(\ref{Act})
becomes quadratic. Therefore, in this
limit, any (even) correlator of $\rho$'s can be evaluated explicitly. In particular,
the bilocal one reads $\left<\rho({\bf x})\rho(0)\right>=-(m_D/g_m)^2
\nabla^2D_{m_D}({\bf x})\simeq 2\xi(1+2\xi I)\delta({\bf x})$, where in the
derivation of the last equality we have used the exponential smallness of $m_D$.
This yields the average squared density: $\overline{\rho^2}={\cal V}^{-1}
\int d^3x\left<\rho({\bf x})\rho(0)\right>\simeq 2\xi{\cal V}^{-1}(1+2\xi I)$.
Next, at $|\rho|\le\xi$,
the average distance between monopoles, $\bar r$, is not smaller than $\xi^{-1/3}$.
The volume of observation, ${\cal V}$, should be much larger than $\bar r^3$
and therefore ${\cal V}$ is much larger than $\xi^{-1}$ as well. This yields the relation
$\overline{\rho^2}\sim\xi{\cal V}^{-1}\ll\xi^2$, which justifies the
initial approximation $|\rho|\ll\xi$.

\section{SU(N)-case}

The $SU(N)$-generalization of the action~(\ref{1}), stemming from the $SU(N)$
Georgi-Glashow model, has the form

\begin{equation}
\label{s}
S=\int d^3x\left[\frac12(\nabla\vec\chi)^2+\frac12(\nabla\psi)^2+
\frac{m_H^2}{2}\psi^2-2\zeta{\rm e}^{g_m\psi}\sum\limits_{i=1}^{N(N-1)/2}
\cos\left(g_m\vec q_i\vec\chi\right)\right].
\end{equation}
Here, $\vec q_i$'s are the positive root vectors of the group $SU(N)$.
As well as the field $\vec\chi$, these vectors are $(N-1)$-dimensional.
Note that the $SU(3)$-version of the action~(\ref{s}), which incorporates
the effects of the Higgs field, has been discussed
in ref.~\cite{nd}. The compact-QED limit of the $SU(N)$-case
has been studied in refs.~\cite{wd}, \cite{sn}, and~\cite{suN}. The string representation
of the compact-QED limit has been studied for the $SU(3)$-case in ref.~\cite{epl} both
in 3D and 4D.
Here, similarly to all the above-mentioned papers, we have assumed that W bosons corresponding to
different root vectors have the same masses.

Straightforward integration over $\psi$ then yields the following analogue of eq.~(\ref{2}):

$$
S\simeq\int d^3x\left[\frac12(\nabla\vec\chi)^2-2\xi\sum\limits_{i=1}^{N(N-1)/2}
\cos\left(g_m\vec q_i\vec\chi\right)\right]-
$$

\begin{equation}
\label{N}
-2\xi^2\int d^3xd^3y\sum\limits_{i,j=1}^{N(N-1)/2}
\cos\left(g_m\vec q_i\vec\chi({\bf x})\right){\cal K}({\bf x}-{\bf y})
\cos\left(g_m\vec q_j\vec\chi({\bf y})\right).
\end{equation}
The Debye mass of the field $\vec\chi$ can be derived from this expression by virtue
of the formula~\cite{group} $\sum\limits_{i=1}^{N(N-1)/2}q_i^\alpha q_i^\beta\propto
\delta^{\alpha\beta}$. The proportionality coefficient which should stand on the
r.h.s. of this relation can easily be found from the requirement that all root vectors have the
unit length. This coefficient is equal to $(N/2)$, and the square of the Debye mass turns out to be
$m_D^2=g_m^2\xi N[1+\xi IN(N-1)]$. Note that this formula reproduces both the $SU(2)$-result
of the previous Section and the $SU(3)$-result of the compact-QED limit~\cite{epl}, \cite{nd}
$m_D^2=3g_m^2\zeta$.

The new parameter of the cumulant expansion, $\xi IN(N-1)$, will be exponentially small
provided that at $x\sim 1/2$,

$$\epsilon(x)>\frac12\left[3{\rm e}^{-\sqrt{2x}}+\frac{g^2}{2\pi m_W}\ln(N(N-1))\right].$$
Setting in this inequality $x=1/2$ and recalling that~\footnote{Similarly to ref.~\cite{suN},
we assume here that the function $\epsilon$ is one and the same for any $N$.}
$\epsilon(1/2)<\epsilon(\infty)\simeq 1.787$,
we obtain the following upper bound on $N$, which guarantees the convergence of the cumulant
expansion: $N(N-1)<{\rm e}^{15.522m_W/g^2}$. Clearly, in the weak-coupling regime under study,
this bound is exponentially large, that allows $N$ to be large enough too.

Next, the representation of the theory with
the action~(\ref{N}) in terms of the monopole densities can be derived
similarly to the $SU(2)$-case by virtue of the formula

$$\exp\left[-\frac12\int d^3x(\nabla\vec \chi)^2\right]=$$

$$=\int\prod\limits_{i=1}^{N(N-1)/2} {\cal D}\rho_i
\exp\left[-\frac{g_m^2}{2}\int d^3xd^3y\rho_i({\bf x})D_0({\bf x}-{\bf y})
\rho_i({\bf y})-ig_m\sqrt{\frac{2}{N}}\int d^3x\vec q_i\vec\chi\rho_i\right].$$
The analogue of the saddle-point equation~(\ref{sp}) then reads

$$
\sinh(\phi_i({\bf x}))\left[1+2\xi\sum\limits_{j=1}^{N(N-1)/2}\int d^3y{\cal K}({\bf x}-{\bf y})
\cosh(\phi_j({\bf y}))\right]=\frac{\rho_i({\bf x})}{\sqrt{2N}\xi},
$$
where $\phi_j\equiv ig_m\vec q_j\vec\chi$. Solving this equation iteratively with the
Ansatz $\phi_i=\phi_i^1+\phi_i^2$, where $|\phi_j^2|\ll |\phi_j^1|$, we obtain:

$$\phi_i^1({\bf x})={\rm arcsinh}\left(\frac{\rho_i({\bf x})}{\sqrt{2N}\xi}\right),~~
\phi_i^2({\bf x})=-\sqrt{\frac{2}{N}}\frac{\rho_i({\bf x})}{F_i({\bf x})}
\sum\limits_{j=1}^{N(N-1)/2}
\int d^3y{\cal K}({\bf x}-{\bf y})F_j({\bf y}),$$
where $F_i({\bf x})\equiv\sqrt{1+\frac{1}{2N}\left(\frac{\rho_i({\bf x})}{\xi}\right)^2}$.
Next, the average density of monopoles of {\it all} kinds reads
$\frac{\partial\ln\int {\cal D}\vec\chi{\rm e}^{-S}}{{\cal V}\partial\ln\xi}
\simeq\xi N(N-1)[1+\xi IN(N-1)]$, that, in particular, reproduces the $SU(2)$-result.
Therefore, the average density of monopoles of only one kind is of the order of $\xi$.
Thus, at $|\rho_i|\le\xi$, $F_i\sim 1$,
$|\phi_i^1|\sim|\rho_i|/(\xi\sqrt{N})$, and
$|\phi_i^2|\sim\xi IN(N-1)|\phi_i^1|$. The quantity $|\phi_i^2|/|\phi_i^1|$ is
therefore of the order of the parameter of the cumulant expansion, that
justifies the adapted Ansatz.
The desired representation of the theory described by
the action~(\ref{N}) in terms of the monopole densities then has the form

$${\cal Z}=\int\left(\prod\limits_{i=1}^{N(N-1)/2} {\cal D}\rho_i\right)\exp\left\{
-\frac{g_m^2}{2}\int d^3xd^3y\sum\limits_{i=1}^{N(N-1)/2}\rho_i({\bf x})D_0({\bf x}-{\bf y})
\rho_i({\bf y})-V_N\left[\{\rho_i\}_{i=1}^{N(N-1)/2}\right]\right\},$$
where the monopole potential is given by the following formula:

$$
V_N\left[\{\rho_i\}_{i=1}^{N(N-1)/2}\right]=\int d^3x\sum\limits_{i=1}^{N(N-1)/2}
\left[\sqrt{\frac{2}{N}}
\rho_i{\,}{\rm arcsinh}\left(\frac{\rho_i}{\sqrt{2N}\xi}\right)
-2\xi F_i\right]-$$

$$-2\xi^2\int d^3xd^3y\sum\limits_{i,j=1}^{N(N-1)/2}
F_i({\bf x}){\cal K}({\bf x}-{\bf y})F_j({\bf y}).$$

One can further naively assume that the criterion of the low-density approximation
has the form $|\rho_i|\ll\sqrt{N}\xi$ (although the average density of monopoles
of one kind was discussed to be of the order of $\xi$). Indeed, similarly to the $SU(2)$-case,
already under this inequality, the potential
factorizes and becomes quadratic, so that (again up to an inessential constant addendum)
$V_N\left[\{\rho_i\}_{i=1}^{N(N-1)/2}\right]\simeq
\frac{g_m^2}{2m_D^2}\int d^3x\sum\limits_{i=1}^{N(N-1)/2}\rho_i^2$.
Consequently, the
bilocal correlator of monopole densities reads

$$\left<\rho_i({\bf x})\rho_j(0)\right>=-(m_D/g_m)^2\delta_{ij}
\nabla^2D_{m_D}({\bf x})\simeq \xi N[1+\xi IN(N-1)]\delta_{ij}\delta({\bf x}),$$
and, in particular,
the $SU(2)$-result obviously recovers itself. Therefore, the average squared
density of monopoles of any kind has the form: $\overline{\rho_i^2}\simeq \xi N{\cal V}^{-1}
[1+\xi IN(N-1)]\sim \xi N{\cal V}^{-1}$. The inequality $\overline{\rho_i^2}\ll N\xi^2$,
necessary for the justification of the initial approximation, will thus
be satisfied provided that ${\cal V}\xi\gg 1$. For the densities $|\rho_i|\le\sqrt{N}\xi$,
we however have ${\cal V}\gg\bar r_i^3\ge N^{-1/2}\xi^{-1}$ (where $\bar r_i$ is an
average distance between the monopoles of the $i$-th kind),
{\it i.e.} ${\cal V}\xi\gg N^{-1/2}$, rather than ${\cal V}\xi\gg 1$.
The initial naive low-density approximation
$|\rho_i|\ll\sqrt{N}\xi$, which ensures the factorization of the potential,
is then fully justified for not too large $N$, {\it i.e.} it should be replaced
by the right one, $|\rho_i|\ll\xi$. In another words,
for too large $N$, the requirement $|\rho_i|\ll\sqrt{N}\xi$ becomes no more
the low-density approximation, since
it then allows $|\rho_i|$ to exceed significantly its
average value, which is of the order of $\xi$.

Note finally that the
obtained results lead to obvious modifications of the values of the confining-string
coupling constants (string tension, coupling constant of the rigidity term, and so on).
These modifications, which are due to the change of the Debye mass of the dual photon,
can be accounted for by virtue of the formulae obtained in ref.~\cite{cu}. One should
also take into account that the charges of quarks are distributed over the lattice of
weight vectors of the group $SU(N)$,
whose squares are equal to $(N-1)/(2N)$. We finally obtain the following values of the
string tension and the inverse coupling constant of the rigidity term ({\it cf.}
refs.~\cite{nd} and~\cite{epl}):

$$\sigma=8\pi^2g\sqrt{\xi}\frac{N-1}{\sqrt{N}}\Biggl[1+\frac12\xi IN(N-1)\Biggr],~~
\alpha^{-1}=-\frac{1}{16}\frac{g^3}{\sqrt{\xi}}\frac{N-1}{N^{3/2}}\Biggl[1-\xi IN(N-1)\Biggr].$$

\section{Conclusions}

In the present paper, we have explored the influence of the Higgs field to the
dynamics of the (2+1)D Georgi-Glashow model and its $SU(N)$-generalization.
To this end, the Higgs field was not supposed to be infinitely heavy, as it
takes place in the compact-QED limit of the model. Owing to this fact, the
Higgs field starts propagating, that leads to the additional interaction
between monopoles and, consequently, to the modification of the conventional
sine-Gordon theory of the dual-photon field. Contrary to the previous analysis,
performed in ref.~\cite{nd} in the BPS limit, in the present paper the Higgs-boson
mass was considered to be of the order of the W-boson mass. In this regime, combined
with the standard weak-coupling approximation, the
Debye mass of the dual photon and the potential of monopole densities have been
found. In the low-density limit, the latter enables one to evaluate correlators
of densities to any order. There has also been demonstrated that the existing data
on the monopole fugacity provide the convergence of the cumulant expansion, which is
used for the average over the Higgs field. This justifies the bilocal approximation
adapted for the performed analysis.

After that, the above-described investigation
has been generalized to the case of the $SU(N)$ Georgi-Glashow model with $N\ge 2$.
The results obtained in this way reproduce, in particular, the respective
$(N=2)$-ones. There has also been found the upper bound for $N$, necessary to
ensure the convergence of the above-mentioned cumulant expansion. This bound is
a certain exponent of the ratio of the W-boson mass to the squared electric
coupling constant. It is therefore an exponentially large quantity in the weak-coupling
regime, that yields an enough broad range for the variation of $N$. Finally, we have found the values
of the two leading coupling constants of the confining-string Lagrangian at arbitrary $N$.

\section{Acknowledgments}
The author is greatful for useful discussions
to Prof. A.~Di~Giacomo and Dr. N.O.~Agasian. He is also greatful to
Prof. A.~Di~Giacomo and to the whole staff of the Physics Department of the
University of Pisa for cordial hospitality.
This work has been supported by INFN and partially by
the INTAS grant Open Call 2000, Project No. 110.

\section*{Appendix. Evaluation of the integral
$\int d^3x{\cal K}({\bf x})$.}

Setting, as everywhere else in this paper, $m_W$ for an UV cutoff
and using the notations for $a$, $c$, $[1/c]$, and the incomplete
Gamma-function introduced in the main text, we have
for the desired integral:

$$I=\frac{4\pi}{m_H^3}\int\limits_{c}^{\infty}dxx^2
\left[\exp\left(\frac{a{\rm e}^{-x}}{x}\right)-1\right]=
\frac{4\pi}{m_H^3}\sum\limits_{n=1}^{\infty}\frac{a^n}{n!}
\int\limits_{c}^{\infty}dx{\rm e}^{-nx}x^{2-n}=$$

$$=\frac{4\pi}{m_H^3}\sum\limits_{n=1}^{\infty}\frac{a^n}{n!}
n^{n-3}\Gamma(3-n,cn)\simeq
\frac{4\pi}{m_H^3}\left[\sum\limits_{n=1}^{[1/c]}\frac{a^n}{n!}
n^{n-3}\Gamma(3-n,cn)+c^2\sum\limits_{[1/c]+1}^{\infty}
\left(\frac{a}{c}\right)^n\frac{{\rm e}^{-cn}}{nn!}\right].\eqno(A.1)$$
Clearly, in the derivation of the last equality, we have used
the asymptotics of the incomplete Gamma-function at large
values of its second argument: $\Gamma(3-n, cn)\simeq (cn)^{2-n}
{\rm e}^{-cn}$.
The last sum in eq.~(A.1) can further be rewritten as
$\sum\limits_{n=1}^{\infty}-\sum\limits_{n=1}^{[1/c]}$, and we obtain:

$$I\simeq\frac{4\pi}{m_H^3}\left\{\sum\limits_{n=1}^{[1/c]}
\frac{a^n}{nn!}\left[n^{n-2}\Gamma(3-n,cn)-c^{2-n}{\rm e}^{-cn}\right]
+c^2\sum\limits_{n=1}^{\infty}\left(\frac{a}{c}\right)^n
\frac{{\rm e}^{-cn}}{nn!}\right\}.\eqno(A.2)$$

Note that the last sum here is equal to
${\rm Ei}\left(\frac{a}{c}{\rm e}^{-c}\right)-\gamma-\ln\left(\frac{a}{c}{\rm
e}^{-c}\right)$,
where $\gamma\simeq 0.577$ is the Euler constant, and Ei denotes the integral
exponential function. However, in the interesting to us case
$c\sim 1$, $a\gg 1$, such a representation of that sum does not
help when one tries to express it explicitly in terms of $a$ and $c$.
Instead, it is useful to rewrite it as follows:

$$\int\limits_{0}^{\infty}dt\sum\limits_{n=1}^{\infty}
\left(\frac{a}{c}\right)^n\frac{{\rm e}^{-(c+t)n}}{n!}=
\int\limits_{0}^{\infty}dt\left\{\exp\left[\frac{a}{c}
{\rm e}^{-(c+t)}\right]-1\right\}=\int\limits_{0}^{\frac{a}{c}
{\rm e}^{-c}}\frac{dz}{z}\left({\rm e}^z-1\right),\eqno(A.3)$$
where $z\equiv\frac{a}{c}{\rm e}^{-(c+t)}$. Integrating
by parts we have at $\frac{a}{c}{\rm e}^{-c}\gg 1$:

$$\left.\left.(A.3)\simeq\left[\exp\left(\frac{a}{c}{\rm e}^{-c}\right)-1\right]
\left(\ln\frac{a}{c}-c\right)-\Biggl(\left<{\rm e}^z\right>\right|_{0}^{1}
\int\limits_{0}^{1}dz\ln z+\left<\ln z\right>\right|_{1}^{\frac{a}{c}{\rm e}^{-c}}
\int\limits_{1}^{\frac{a}{c}{\rm e}^{-c}}dz{\rm e}^z\Biggr)\simeq$$

$$
\simeq\exp\left(\frac{a}{c}{\rm e}^{-c}\right)\left(1-\frac{c}{a}{\rm e}^c\right)+
({\rm e}-1)\ln a,$$
where in the derivation of the last equality we have kept the terms leading in $a$ and $(a/c)$.
Together with the first sum standing on the r.h.s. of
eq.~(A.2) this
finally yields eq.~(\ref{3}) of the main text.

\end{document}